\documentclass{PoS}

\usepackage{amsmath}
\usepackage{amssymb}
\usepackage{graphicx}

\newcommand{\be}{\begin{equation}}
\newcommand{\ee}{\end{equation}}
\newcommand{\ba}{\begin{align}}
\newcommand{\ea}{\end{align}}

\newcommand{\chibar}{{\bar{\chi}}}

\title{Aoki Phases in Staggered-Wilson Fermions\thanks{YITP-11-86, KUNS-2366, RIKEN-MP-33}}

\ShortTitle{Aoki Phases in Staggered-Wilson Fermions}

\author{\speaker{Tatsuhiro Misumi}\\ 
        Yukawa Institute for Theoretical Physics, Kyoto University\\\
        E-mail: \email{misumi@yukawa.kyoto-u.ac.jp}}

\author{Michael Creutz \thanks{Authored under contract number
DE-AC02-98CH10886 with the U.S.~Department of Energy.  Accordingly,
the U.S. Government retains a non-exclusive, royalty-free license to
publish or reproduce the published form of this contribution, or allow
others to do so, for U.S.~Government purposes.}\\
        Brookhaven National Laboratory\\
        E-mail: \email{creutz@bnl.gov}}
        
\author{Taro Kimura\\
        Department of Basic Science, University of Tokyo\\
        Mathematical Physics Laboratory, RIKEN\\
        E-mail: \email{kimura@dice.c.u-tokyo.ac.jp}}
        
\author{Takashi Z. Nakano\\
        Department of Physics, Kyoto University\\
        Yukawa Institute for Theoretical Physics, Kyoto University\\
        E-mail: \email{tnakano@yukawa.kyoto-u.ac.jp}}
        
\author{Akira Ohnishi\\
        Yukawa Institute for Theoretical Physics, Kyoto University\\
        E-mail: \email{ohnishi@yukawa.kyoto-u.ac.jp}}

\abstract{We investigate the parity-broken phase (Aoki phase) 
for staggered-Wilson fermions by using the Gross-Neveu model and the 
strong-coupling lattice QCD. In the both cases the gap equations indicate 
the parity-broken phase exists and the pion becomes massless on the phase
boundaries. We also show we can take the chiral and continuum limit 
in the Gross-Neveu model by tuning mass and gauge-coupling parameters. 
This supports the idea that the staggered-Wilson fermions can be applied to the
lattice QCD simulation by taking a chiral limit, as with Wilson fermions.
}

\FullConference{ The XXIX International Symposium on Lattice Field Theory - Lattice 2011\\
July 10-16, 2011\\
Squaw Valley, Lake Tahoe, California}

\begin{document}


\section{Introduction}
\label{sec:Intro}

Recently staggered-based Wilson fermions were proposed 
by introducing the taste-splitting mass or the flavored-mass terms into
staggered fermions \cite{Adams, Hoel, CKM2}. They can be 
applied to lattice QCD not only as Wilson fermions but also as an overlap 
kernel. One possible advantage of these novel fermions called
staggered-Wilson and staggered-overlap is reduction of the matrix 
sizes in the associated Dirac operators, which leads to reduction of 
numerical costs in lattice QCD simulations. Thus they may be able to overcome the 
usual naive-fermion-based lattice fermions in lattice QCD \cite{PdF}. 
The purpose of this work is reveal properties of staggered Wilson fermions
in terms of the parity phase structure (Aoki phase) \cite{AokiP}. 
The Aoki phase for the staggered-Wilson was first studied in Ref.~\cite{CKM2}
and the present paper shows further investigation of this topic.
The existence of the Aoki phase and the second-order phase boundary in 
Wilson-type lattice fermions indicates that one can apply them to lattice 
QCD simulations by tuning a mass parameter to take a chiral limit. Besides, 
the understanding of the parity-broken phase gives practical information 
for the application of its overlap and domain-wall versions. 

In this paper we elucidate the parity phase structure for staggered-Wilson
fermions in the framework of the Gross-Neveu model and the hopping 
parameter expansion in the strong-coupling lattice QCD. We find the gap 
equations derived from the both theories show the pion condensate
becomes nonzero in some range of the parameters and the pion 
becomes massless on the phase boundaries. It means the Aoki phase
exists and the order of the phase transition is second-order.
We also show we can take the chiral continuum limit 
in the Gross-Neveu model by tuning the mass and the gauge-coupling. 
These results on the staggered-Wilson fermion 
incidate we can obtain one- or two-flavor fermions by tuning 
the mass parameter and perform the lattice QCD simulation with 
these fermions as in the Wilson fermion. We note the results on the 
Gross-Neveu model is based on the work by some of the present authors 
\cite{CKM2, CKM1} while the results on the strong-coupling lattice QCD 
are parts of a work in progress.


\section{Staggered Wilson fermions}
\label{sec:SWF}
We begin with staggered-Wilson fermions in which the flavored-mass terms 
split the four degenerate tastes in a manner similar to the usual Wilson term.
There are two possible types of the flavored-mass terms for staggered
fermions as
\begin{align}  
M_{f}^{(1)} &= \epsilon\sum_{sym} \eta_{1}\eta_{2}\eta_{3}\eta_{4}
C_{1}C_{2}C_{3}C_{4}
= ({\bf 1} \otimes \gamma_{5}) + O(a),
\\
M_{f}^{(2)} &=\sum_{\mu>\nu}{i\over{2\sqrt{3}}}\epsilon_{\mu\nu}
\eta_{\mu}\eta_{\nu}(C_{\mu}C_{\nu}+C_{\nu}C_{\mu}) 
= ({\bf 1} \otimes \sum_{\mu>\nu}\sigma_{\mu\nu}) + O(a),
\end{align}
where $C_{\mu}=(V_{\mu}+V_{\mu}^{\dag})/2$,  
$(\eta_{\mu})_{xy}=(-1)^{x_{1}+...+x_{\mu-1}}\delta_{x,y}$,
$(\epsilon)_{xy}=(-1)^{x_{1}+...+x_{4}}\delta_{x,y}$,
$(\epsilon_{\mu\nu})_{xy}=(-1)^{x_{\mu}+x_{\nu}}\delta_{x,y}$, 
with $(V_{\mu})_{xy}=U_{\mu,x}\delta_{y,x+\mu}$. 
In the right hand sides we use the spin-taste representation as ${\bf 1}\otimes \gamma_{5}$. 
We refer to $M_{f}^{(1)}$ as the Adams-type and $M_{f}^{(2)}$ the Hoelbling-type. 
The former splits the 4 tastes into two with positive($m=+1$) and the other two 
with negative($m=-1$) mass while the latter split them into one with positive($m=+2$), 
two with zero($m=0$) and the other one with negative mass($m=-2$).
Now we introduce the Wilson parameter $r=r\delta_{x,y}$ and shift the mass for the actions
as with Wilson fermions. Then the Adams-type staggered-Wilson fermion action 
is given by
\begin{equation}  
S_{\rm A}\,=\,\sum_{xy}\bar{\chi}_{x}[\eta_{\mu}D_{\mu}+r(1+M_{f}^{(1)})+M]_{xy}\chi_{y},
\end{equation}
with $D_{\mu}={1\over{2}}(V_{\mu}-V_{-\mu})$.
Here $M$ stands for the usual taste-singlet mass ($M=M\delta_{x,y}$). 
The Hoelbling-type staggered-Wilson fermion action is given by
\begin{align}  
S_{\rm H}\,=\,\sum_{xy}\bar{\chi}_{x}[\eta_{\mu}D_{\mu}+r(2+M_{f}^{(2)})+M]_{xy}\chi_{y}.
\label{HoelS}
\end{align}
In the QCD simulation we will tune the mass parameter $M$ to take a 
chiral limit. For some negative values of the mass parameter:$-1<M<0$ 
for Adams-type and $-2<M<0$ for Hoelbling-type with $r=1$, we obtain two-flavor 
and one-flavor overlap fermions respectively by using the overlap formula.

The potential problem in lattice QCD with these fermions is the breaking of
some discrete symmetries as the shift symmetry caused by the flavored-mass 
terms \cite{Adams, Hoel}. There has not yet been a consensus on 
whether it does harm to lattice QCD with staggered-Wilson fermions. We 
can answer this question partly by studying the Aoki phase since a clear 
symptom is expected to appear in the phase structure if the symmetry 
breaking ruins the essential properties of QCD. In the following sections 
we will find the Aoki phase structure in the staggered-Wilson fermion is 
qualitatively similar to the original Wilson one and there is no disease.


\section{Gross-Neveu model}
\label{sec:GN}

\begin{figure}
\centering
\includegraphics[height=4cm]{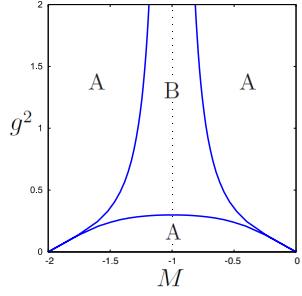} 
\caption{Aoki phase structure for the staggered-Wislon fermion
in the Gross-Neveu model.  A stands 
for a parity symmetric phase and B for Aoki phase.}
\label{fig4}
\end{figure}

We first investigate the parity phase diagram for staggered-Wilson fermions 
by using the $d=2$ Gross-Neveu model as a toy model of QCD.
To study the pion condensate we generalize the usual staggered Gross-Neveu model 
to the one with the $\gamma_{5}$-type 4-point interaction, which is given by
\begin{align}
S\,=\, {1\over{2}}&\sum_{n,\mu}\eta_{\mu}\bar{\chi}_{n}(\chi_{n+\mu}-\chi_{n-\mu})
+\sum_{n}\bar{\chi}_{n}(M+r(1+M_{f}))\chi_{n}
\nonumber\\
&-{g^{2}\over{2N}}\sum_{\mathcal{N}}\Big[(\sum_{A}\bar{\chi}_{2\mathcal{N}+A}\,\chi_{2\mathcal{N}+A})^{2}+(\sum_{A}i(-1)^{A_{1}
+A_{2}}\bar{\chi}_{2\mathcal{N}+A}\,\chi_{2\mathcal{N}+A})^{2}\Big],
\label{Ssta}
\end{align}
where the two-dimensional coordinate is defined as $n=2\mathcal{N}+A$ with sublattices $A=(A_{1},A_{2})$($A_{1,2}=0,1$).  In this model $\chi_{n}$ is a $N$-component one-spinor
$(\chi_{n})_{j}$($j=1,2,...,N$) where $\bar{\chi}\chi=\sum_{j=1}^{N}\bar{\chi}_{j}\chi_{j}$.
$(-1)^{A_{1}+A_{2}}$ corresponds to $\Gamma_{55}=\gamma_{5}\otimes\gamma_{5}$ 
in the spinor-taste expression while $\eta_{\mu}=(-1)^{n_{1}+...+n_{\mu-1}}$ corresponds to $\gamma_{\mu}$. 
In this dimension the Adams-type and Hoelbling-type flavored-mass terms coincide and 
there is only one type $M_{f}=\Gamma_{5}\Gamma_{55}\sim{\bf 1}\otimes\gamma_{5}+O(a)$
with $\Gamma_{5}=-i\eta_{1}\eta_{2}\sum_{\rm sym}C_{1}C_{2}$.
This mass term assigns the positive mass ($m=+1$) to one taste and the negative mass ($m=-1$) to the other.
With bosonic auxiliary fields $\sigma_{\mathcal{N}}$, $\pi_{\mathcal{N}}$ leading to $\sigma$-meson and $\pi$-meson fields, the action is rewritten as
\begin{align}
S\,=\, {1\over{2}}\sum_{n,\mu}&\eta_{\mu}\bar{\chi}_{n}(\chi_{n+\mu}-\chi_{n-\mu})
+\sum_{n}\bar{\chi}_{n}M_{f}\chi_{n}
\nonumber\\
&+{N\over{2g^{2}}}\sum_{\mathcal{N}}((\sigma_{\mathcal{N}}-1-M)^{2}+\pi_{\mathcal{N}}^{2})
+\sum_{\mathcal{N},A}\bar{\chi}_{2\mathcal{N}+A}(\sigma_{\mathcal{N}}+i(-1)^{A_{1}+A_{2}}\pi_{\mathcal{N}})\chi_{2\mathcal{N}+A},
\label{StS}
\end{align}
where we take $r=1$ as the Wilson parameter.
After integrating the fermion field, the partition function and the effective action 
with these auxiliary fields(meson fields) are given by
\begin{align}
Z= \int \mathcal{D}\sigma_{\mathcal{N}}\mathcal{D}\pi_{\mathcal{N}}e^{-N\,S_{\rm eff}(\sigma,\pi)},
\,\,\,\,\,\,\,\,\,\,\,\,\,\,\,\,\,\,\,\,\,\,\,
S_{\rm eff} = {1\over{2g^{2}}}\sum_{\mathcal{N}}((\sigma_{\mathcal{N}}-1-M)^{2}+\pi_{\mathcal{N}}^{2})-{\rm Tr}\,\log D,
\label{StSeff}
\end{align}
with $D_{n,m}=(\sigma_{\mathcal{N}}+i(-1)^{A_{1}+A_{2}}\pi_{\mathcal{N}})\delta_{n,m}+
{\eta_{\mu}\over{2}}(\delta_{n+\mu, m}-\delta_{n-\mu, m})+(M_{f})_{n,m}$.
In the large N limit the partition function is given by the saddle point of the action as
$Z\,=\, e^{-S_{\rm eff}(\sigma_{0},\pi_{0})}$ with the translation-invariant solutions $\sigma_{0}$, $\pi_{0}$ satisfying the saddle-point equations ${\delta S_{\rm eff}(\sigma_{0},\pi_{0})\over{\delta \sigma_{0}}}={\delta S_{\rm eff}(\sigma_{0},\pi_{0})\over{\delta \pi_{0}}}=0$.
After some calculation process to derive the fermion determinant \cite{CKM2}
we obtain the concrete forms of the saddle-point equations in the momentum space
\begin{align}
{\sigma_{0}-1-M\over{g^{2}}}&=4\int{dk^{2}\over{(2\pi)^2}}{\sigma_{0}(\sigma_{0}^{2}+\pi_{0}^{2}+s^{2})-c_{1}^{2}c_{2}^{2}\sigma_{0}\over{((\sigma_{0}+c_{1}c_{2})^{2}+\pi_{0}^2+s^{2})((\sigma_{0}-c_{1}c_{2})^2+\pi_{0}^2+s^{2})}},
\label{sad1}
\\
{\pi_{0}\over{g^{2}}}&=4\int{dk^{2}\over{(2\pi)^2}}{\pi_{0}(\sigma_{0}^{2}+\pi_{0}^{2}+s^{2})
+c_{1}^{2}c_{2}^{2}\pi_{0}\over{((\sigma_{0}+c_{1}c_{2})^{2}+\pi_{0}^2+s^{2})((\sigma_{0}-c_{1}c_{2})^2+\pi_{0}^2+s^{2})}},
\label{sad2}
\end{align}
with $c_{\mu}=\cos k_{\mu}/2$ and $s_{\mu}=\sin k_{\mu}/2$.
Now what we are interested in is the parity phase diagram in this theory.
The parity phase boundary $M_{c}(g^{2})$ is derived by imposing 
$\pi_{0}=0$ in (\ref{sad1})(\ref{sad2}) after the overall $\pi_{0}$ being removed in the second one.
Then the gap equations are given by
\begin{align} 
{1+M_{c}\over{g^{2}}}&=4\int{dk^{2}\over{(2\pi)^2}}{2c_{1}^{2}c_{2}^{2}\sigma_{0}
\over{((\sigma_{0}+c_{1}c_{2})^{2}+\pi_{0}^2+s^{2})((\sigma_{0}-c_{1}c_{2})^2+\pi_{0}^2+s^{2})}},
\\
{1\over{g^{2}}}&=4\int{dk^{2}\over{(2\pi)^2}}{\sigma_{0}^{2}+s^{2}+c_{1}^{2}c_{2}^{2}
\over{((\sigma_{0}+c_{1}c_{2})^{2}+\pi_{0}^2+s^{2})((\sigma_{0}-c_{1}c_{2})^2+\pi_{0}^2+s^{2})}}.
\end{align}
By removing $\sigma_{0}$ in these equations, we derive the phase boundary $M_{c}(g^{2})$.
The result is shown in Fig.~\ref{fig4}.
It indicates the parity phase structure in the staggered-Wilson fermion is 
qualitatively similar to the usual Wilson case \cite{AokiP} reflecting the mass splitting 
of tastes given by the flavored mass. We also check the pion mass becomes 
zero on the second order phase boundary as
$m_{\pi}^{2}\, \propto\, 
V{\delta^{2}\tilde{S}_{\rm eff}\over{\delta^{2}\pi_{0}^{2}}}|_{M=M_{c}}=0$
where $S_{\rm eff}=V{\tilde S_{\rm eff}}$ with $V$ being the volume.

We next consider the chiral and continuum limit of the
staggered-Wilson Gross-Neveu models.
The strategy is to expand the fermion 
determinant in the effective potential in Eq.~(\ref{StSeff}) with respect 
to the lattice spacing $a$.
After some calculations (See details in Ref.~\cite{CKM2}) we obtain the effective 
potential remaining in the limit $a\to 0$,
\begin{align}
\tilde{S}_{\rm eff}&=-\Big({M+1/a\over{g_{\sigma}^2}}+{2\over{a}}C_{1}\Big)\sigma_{0}
+\Big( {1\over{2g_{\pi}^2}}-\tilde{C}_{0} + {1\over{\pi}}\log 4a^{2} \Big)\pi_{0}^{2}
\nonumber\\
&+\Big( {1\over{2g_{\sigma}^2}}-\tilde{C}_{0}+ 2C_{2} + {1\over{\pi}}\log 4a^{2} \Big)\sigma_{0}^{2}
+{1\over{\pi}}(\sigma_{0}^{2}+\pi_{0}^2)\log{\sigma_{0}^{2}+\pi_{0}^2\over{e}}.
\label{renSst}
\end{align}
with the three numbers as $\tilde{C}_{0}=1.177$, $C_{1}=-0.896$ and $C_{2}=0.404$.
Here taking the chiral limit means restoring the rotational symmetry in $\sigma_{0}$ 
and $\pi_{0}$ by tuning the parameters.
In this model we need introduce two independent coupling constants 
$g_{\sigma}^{2}$ and $g_{\pi}^{2}$ to restore the symmetry 
although the necessity of two couplings is just a model artifact. 
The tuned point for the chiral limit without $O(a)$ corrections is
\begin{align}
M=-{2g_{\sigma}^{2}\over{a}}C_{1}-1,
\,\,\,\,\,\,\,\,\,\,\,\,\,\,\,
g_{\pi}^{2}={g_{\sigma}^{2}\over{4C_{2}g_{\sigma}^{2}+1}},
\label{gC2st}
\end{align}
To take the continuum limit we introduce the $\Lambda$-parameter as
$2a\Lambda = {\rm exp}\left[{\pi\over{2}} \tilde{C}_{0}-\pi C_{2}-{\pi\over{4g_{\sigma}^{2}}}\right]$.
Then the coupling renormalization for the chiral and continuum limit is given by
\begin{align}
{1\over{2g_{\sigma}^{2}}}=\tilde{C}_{0}- 2C_{2} + {1\over{\pi}}\log \left( {1\over{4\Lambda^{2}a^{2}}}\right),
\,\,\,\,\,\,\,\,\,\,\,\,\,\,
{1\over{2g_{\pi}^{2}}}=\tilde{C}_{0} + {1\over{\pi}}\log \left( {1\over{4\Lambda^{2}a^{2}}}\right).
\label{arengpst}
\end{align}
where we keep $\Lambda$ finite when taking the continuum limit $a\to 0$.
Finally the renormalized effective potential in the chiral and continuum limit is given by
\begin{equation}
\tilde{S}_{\rm eff}={1\over{\pi}}(\sigma_{0}^{2}+\pi_{0}^{2})\log{\sigma_{0}^{2}+\pi_{0}^{2}\over{e\Lambda^{2}}},
\label{renaSst}
\end{equation}
This wine-bottle potential yields the spontaneous breaking of the rotational symmetry.
We have shown that the chirally-symmetric continuum limit can be taken by fine-tuning 
a mass parameter and two coupling constants in the staggered-Wilson Gross-Neveu model. 
Considering that the necessity of the two coupling constants is just a model artifact, 
this result indicates we can take a chiral limit by tuning only the mass parameter as in 
the Wilson fermion. Our results on the chiral and continuum limit for staggered-Wilson are 
qualitatively the same as the Wilson case \cite{AokiP}.


\section{Strong-coupling QCD}

\begin{figure}
 \begin{center}
  \includegraphics[height=2cm]{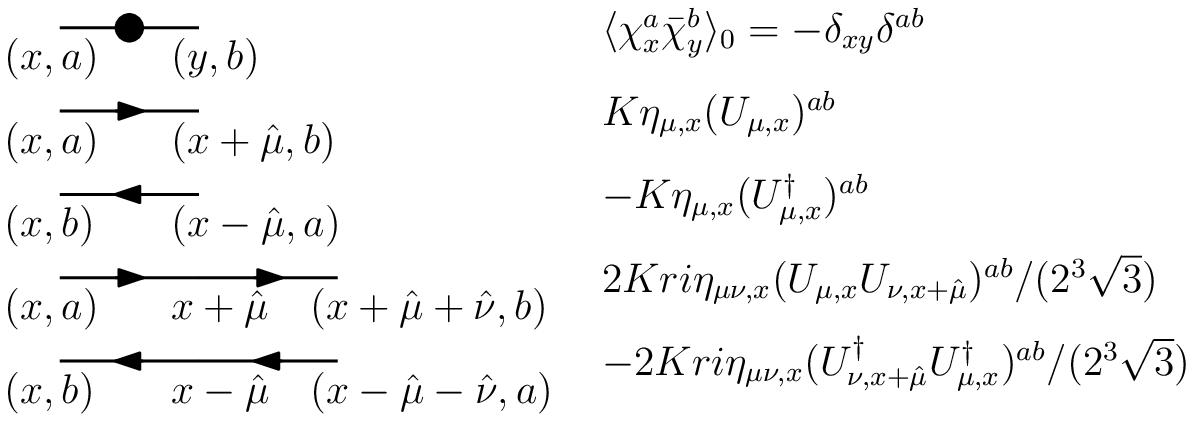}\,\,\,\,
  \includegraphics[height=1.5cm]{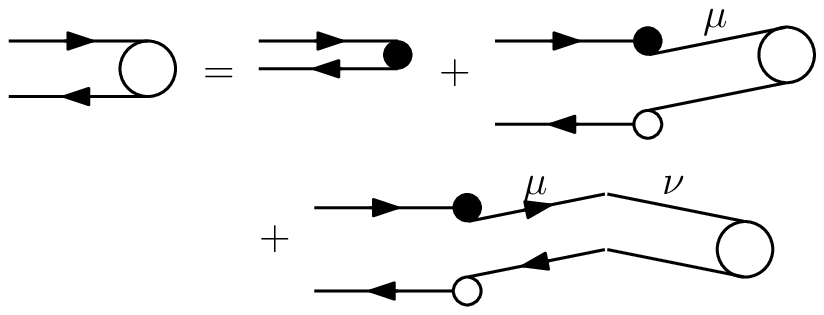}\,\,\,\,\,\,\,\,\,
  \includegraphics[height=2cm]{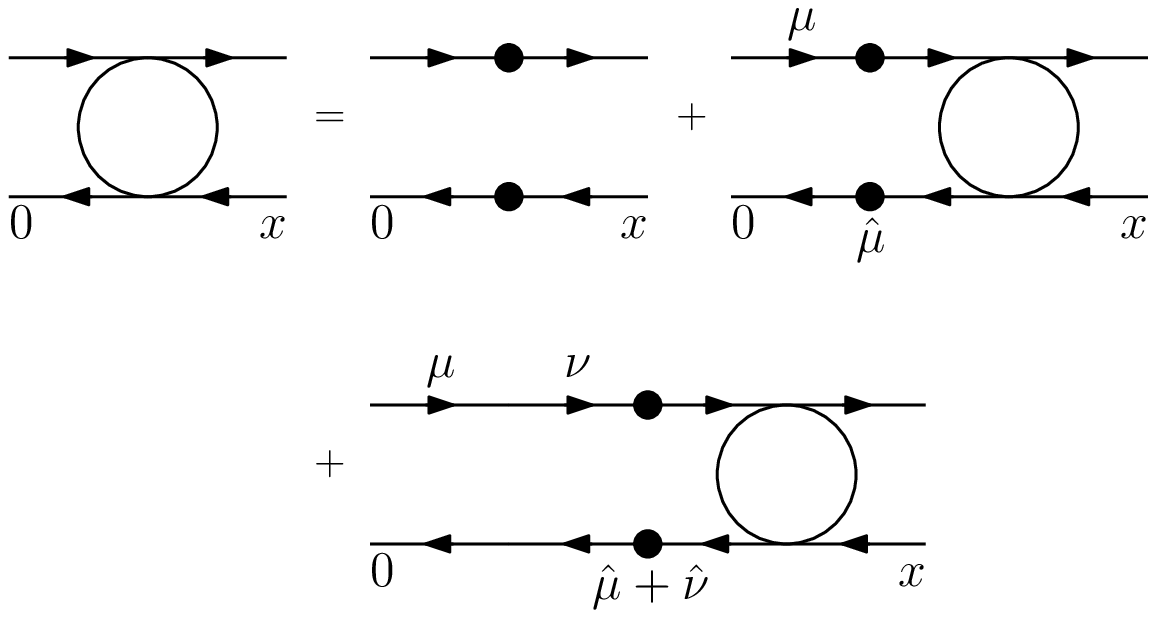}
 \end{center}
 \caption{(Left):Feynman rules for the HPE.
 (Center):one point function. (Right):two point function.}
 \label{FR}
\end{figure}

In this section we investigate the Aoki phase structure in lattice QCD with 
staggered-Wilson fermions in the framework of the hopping parameter 
expansion (HPE) in the strong-coupling regime. 
We can detect a symptom of symmetry breaking from this analysis
although we cannot know details of the true vacuum from this. 
For simplicity we concentrate on the Hoelbling-type lattice fermion here, 
but we can also make the same analysis in a parallel way for the Adams-type fermion.
To perform the HPE for the Hoelbling-type fermion, we rewrite the action
(\ref{HoelS}) by redefining $\chi \rightarrow \sqrt{2K} \chi$ with $K=1/[2(M+2r)]$,
\begin{align}
S &=
\sum_{x} \chibar_x \chi_x + 2K \sum_{x, y} \chibar_{x}(\eta_{\mu}  D_\mu )_{xy}
\chi_{y}+ 2 K r \sum_{x,y} \chibar_x (M_f)_{xy} \chi_y.
\end{align}
In Fig.~\ref{FR}(Left) we write down the Feynman
rules in the HPE for this fermion.
In this paper we perform the hopping parameter 
expansion up to $O(K^{3})$, which works for a small $K$.
We derive chiral and pion condensates from the one-point
function of the meson operator ($\mathcal{M}_{x}=\bar{\chi}_{x}\chi_{x}$) 
in the mean-field approximation. The equation for the one-point function 
up to $O(K^{3})$ is obtained in a self-consistent way as shown in Fig. \ref{FR}(Center),
\begin{equation}
- \Sigma_x  \equiv \langle \mathcal{M}_x \rangle =
 \langle \mathcal{M}_x \rangle_0 + 2K^2 \sum_\mu \Sigma_{x+\hat{\mu}} \Sigma_x
 - 2 \cdot \displaystyle \frac {1}{24} (Kr)^2 \sum_{\mu \neq \nu} \Sigma_x 
 \Sigma_{x+\hat{\mu}+\hat{\nu}},
\label{Eq:HPE-Hoelbling}
\end{equation}
where we drop the link variable since we work in the strong-coupling limit.
$O(K^{3})$ diagrams are found to vanish due to
the cancellation between diagrams.
Here we solve this equation for the condensate $\Sigma$
within the mean-field approximation.
For our purpose we assume $\Sigma_x=\sigma_x +i \epsilon_x \pi_x$,
where $\sigma_x$ and $\pi_x$ correspond to the chiral and pion condensates.
We substitute this form into Eq. (\ref{Eq:HPE-Hoelbling}) and obtain the 
self-consistent equation for $\sigma_{x}$ and $\pi_{x}$ as
\begin{equation}
- \left( \sigma +i \epsilon_x \pi \right)=-1 + 2K^2 \cdot 4 \left( \sigma^2 
+ \pi^2 \right) -2 \cdot \displaystyle \frac {1}{24} (Kr)^2 \cdot 4 \cdot 3 \left( \sigma 
+i \epsilon_x \pi \right)^2,
\end{equation}
which yields $- \sigma = -1 + 16 K^2 \pi^2$ and $- i \pi = - 8K^2 \cdot 2 i \sigma \pi$.
Here we have set $r=2\sqrt{2}$ for simplicity.
We have two solutions depending on whether $\pi=0$ or $\pi\not=0$:
For $\pi=0$ we have a trivial solution $\sigma=1$.
For $\pi \neq 0$ we have a non-trivial solution as
\begin{equation}
\sigma = \displaystyle \frac{1}{16K^2},\,\,\,\,\,\,\,\,\,\,\,\,\,\,
\pi = \pm \sqrt{ \displaystyle \frac{1}{16K^2} \left( 1- \displaystyle \frac{1}{16K^2} \right) }.
\label{cond}
\end{equation}
In this solution the pion condensate is non-zero and the $\pm$ signs 
indicate spontaneous parity breaking.
This parity-broken phase (Aoki phase) appears in the range of the hopping parameter
or the mass parameter as $\mid K \mid > 1/4$ or equivalently 
$-4\sqrt{2}-2<M<-4\sqrt{2}+2$.
We expect that the expansion up to $O(K^{3})$ works to give a meaningful result 
at least around the critical parameter $|K_{c}|^{3}=1/64\ll 1$. 

We next discuss the two-point function of the meson operator $\mathcal{S}(0,x)\equiv\mathcal{M}_{0}\mathcal{M}_{x}$ up to the same order as the one-point function. 
From Fig. \ref{FR}(Right) we derive the following equation for two point function.
($O(K^{3})$ diagrams vanish again.)
\begin{align}
\mathcal{S}(0,x)  \equiv
 \langle \chibar_0^a\chi_0^a \chibar_x^b \chi_x^b \rangle
 = - \delta_{0x} N_c &+ K^2 
  \sum_{\pm \mu} \langle \chibar_{\hat{\mu}}^a \chi_{\hat{\mu}}^a \chibar_x^b \chi_x^b \rangle
  \nonumber\\
  &+ \left( 2 K r i \displaystyle \frac{1}{2^3 \sqrt{3}} \right)^2
  \sum_{\substack{\pm \mu,\pm \nu \\ (\mu \neq \nu)}} 
  \langle \chibar_{\hat{\mu}+\hat{\nu}}^a \chi_{\hat{\mu}+\hat{\nu}}^a \chibar_x^b \chi_x^b \rangle .
\end{align}
Then the self-consistent equation for $\mathcal{S}$ is given in the momentum space as
\begin{align}
\mathcal{S}(p) =
 - N_c + 
 \biggl[
  - K^2 \sum_\mu \left( e^{-ip_\mu} + e^{ip_\mu} \right)
  + \left( 2 K r \displaystyle \frac{1}{2^3 \sqrt{3}} \right)^2
  \sum_{\mu \neq \nu} \sum_{\pm}e^{i( \pm p_\mu \pm  p_\nu)}
 \biggr] \mathcal{S}(p).
\end{align}
We finally obtain the meson propagator as
\begin{equation}
\mathcal{S}(p) = 
N_c
\biggl[ - 2 K^2 \sum_\mu \cos p_\mu + 4 \left( 2 K r \displaystyle \frac{1}{2^3 \sqrt{3}} \right)^2
 \sum_{\mu \neq \nu} \cos p_\mu \cos p_\nu - 1 \biggr]^{-1}.
\label{MP}
\end{equation}
The pole of $\mathcal{S}(p)$ gives the meson mass.
Remembering $\gamma_{5}$ in the staggered fermion is given by
$\epsilon_{x}=(-1)^{x_{1}+...+x_{4}}$ and the pion operator is given by
$\pi_{x}=\bar{\chi}_{x}i\epsilon_{x}\chi_{x}$, it is obvious that the momentum of
the pion should be measured from the shifted origin $p=(\pi,\pi,\pi,\pi)$.
Thus we set $p=(i m_\pi a + \pi, \pi,\pi,\pi)$ for 
$1/\mathcal{S}(p)=0$ in (\ref{MP}), which gives the pion mass $m_\pi$ as
$\cosh(m_\pi a) = 1 + \displaystyle \frac{1-16K^2}{6K^2}$.
In this result the pion mass becomes tachyonic in the range 
$\mid K \mid > 1/4$. It indicates there occurs a phase transition between
parity-symmetric and broken phases at $|K_{c}|=1/4$,
which is consistent with the result of condensates in Eq.~(\ref{cond}).

In this section we investigate parity phase structure for staggered-Wilson fermions
by using the hopping parameter expansion up to $O(K^{3})$ in the strong-coupling regime. 
Although we cannot give a strong argument just from this approximate calculation, 
we note that our result on the phase structure is qualitatively consistent with that of HPE 
for Wilson fermions \cite{AokiP}, which implies existence of the Aoki phase in 
staggered-Wilson fermions. The full-order calculation in HPE and the effective 
potential analysis in progress \cite{NMKO} will give us more conclusive evidence for this topic.


\section{Summary}
\label{sec:S}

In this paper we study the Gross-Neveu model and the strong-coupling lattice 
QCD with staggered Wilson fermions with emphasis on the Aoki phase structure. 
We have shown the parity broken phase and the second order 
phase boundary exist in the staggered-Wilson fermions as with the Wilson fermion.
Our results indicate that we can apply the staggered Wilson fermions
to lattice QCD simulations by mass parameter tuning. 
These results also indirectly suggest the applicability 
of the staggered overlap and staggered domain-wall fermions to lattice QCD. 
We note our results on the Aoki phase diagram exhibit no diseases due to 
a discrete symmetry breaking, which is consistent with the results in the 
lattice perturbation in \cite{Adams, Hoel}.


\begin{thebibliography}{99}


\bibitem{Adams}
D.~H.~Adams, Phys. Rev. Lett. {\bf 104}, 141602 (2010): 
Phys. Lett. B {\bf 699}, 394 (2011).

\bibitem{Hoel}
C.~Hoelbling, Phys. Lett. B {\bf 696}, 422 (2011) [arXiv:1009.5362].

\bibitem{CKM2}
M.~Creutz, T.~Kimura and T.~Misumi,  Phys. Rev. D {\bf 83}, 094506 (2011)
[arXiv:1101.4239].

\bibitem{PdF}
P. de Forcrand, A. Kurkela and M. Panero, PoS Lattice2010 (2011) 080,
[arXiv:1102.1000].

\bibitem{AokiP}
S.~Aoki, Phys. Rev. D {\bf 30}, 2653 (1984);
S. Aoki and K. Higashijima, Prog. Theor. Phys. {\bf 76}, 521.

\bibitem{CKM1}
M.~Creutz, T.~Kimura and T.~Misumi, JHEP 1012, 041 (2010) 
[arXiv:1011.0761].

 \bibitem{NMKO}
T.~Z.~Nakano, T.~Misumi, T.~Kimura and A.~Ohnishi, work in progress.



\end{thebibliography}
\end{document}